\documentclass[usenatbib]{mnras}

\pdfoutput=1
\usepackage{amssymb,amsmath}
\usepackage{graphicx}
\usepackage{multirow}
\usepackage{natbib}
\usepackage{tabularx}
\usepackage{footmisc}
\usepackage{wasysym}
\usepackage{upgreek}
\usepackage{caption}
\usepackage{footnote}

\usepackage{color}
\title{Are the distributions of Fast Radio Burst properties consistent with a cosmological population? }

\author[M.~Caleb et al.]
{M.~Caleb$^{1,2,3}$\thanks{Email: manisha.caleb@anu.edu.au},
C.~Flynn$^{2,3}$,
M.~Bailes$^{2,3}$,
E.D.~Barr$^{2,3}$,
R.W.~Hunstead$^{4}$
E.F.~Keane$^{5,2,3}$, 
\newauthor  V.~Ravi$^{2}$,
W.~van Straten$^{2}$
\\ \\
$^{1}$ Research School of Astronomy and Astrophysics, Australian National University, ACT, 2611, Australia\\
$^{2}$ Centre for Astrophysics and Supercomputing, Swinburne University of Technology, P.O. Box 218, 
Hawthorn, VIC 3122, Australia \\
$^{3}$ ARC Centre of Excellence for All-sky Astrophysics (CAASTRO)\\
$^{4}$ Sydney Institute for Astronomy (SIfA), School of Physics, The University of Sydney, NSW 2006, Australia\\
$^{5}$ SKA Organisation, Jodrell Bank Observatory, Cheshire, SK11 9DL, UK}

\begin{document}


\maketitle

\label{firstpage}


\begin{abstract}
\noindent High time resolution radio surveys over the last few years
have discovered a population of millisecond-duration transient bursts
called Fast Radio Bursts (FRBs), which remain of unknown origin. FRBs
exhibit dispersion consistent with propagation through a cold plasma
and dispersion measures indicative of an origin at cosmological
distances. In this paper we perform Monte Carlo simulations of a
cosmological population of FRBs, based on assumptions consistent with
observations of their energy distribution, their spatial density as a
function of redshift and the properties of the interstellar and
intergalactic media. We examine whether the dispersion measures,
fluences, derived redshifts, signal-to-noise ratios and effective widths of known FRBs are consistent with a
cosmological population. Statistical analyses indicate that at least
50 events at Parkes are required to distinguish between a constant
comoving FRB density, and a FRB density that evolves with redshift
like the cosmological star formation rate density.

\end{abstract}



\section{Introduction}

Fast Radio Bursts (FRBs) are bright (few Jy), radio pulses occurring
with time-scales of order milliseconds. Eighteen bursts have been
discovered to date (\citeauthor{Lorimer} \citeyear{Lorimer};
\citeauthor{Thornton} \citeyear{Thornton}; \citeauthor{Spitler}
\citeyear{Spitler}; \citeauthor{Spolaor-Bannister}
\citeyear{Spolaor-Bannister}; \citeauthor{Petroff_FRB}
\citeyear{Petroff_FRB}; \citeauthor{Ravi} \citeyear{Ravi}; 
\citeauthor{Champion} \citeyear{Champion}; \citeauthor{Masui} \citeyear{Masui};
 Keane et al. in prep; Ravi et al. in prep) The integrated electron densities
along the lines of sight to these bursts (called dispersion measures,
or DMs) lie in the range of 375 to 1600 $\mathrm{pc\,cm^{-3}}$. This
is greatly in excess of the expected contribution from the Galaxy via
the Interstellar Medium (ISM) \citep{Cordes} along such
lines-of-sight, which typically lie in the range of 20 to 50
$\mathrm{pc\,cm^{-3}}$.


For all FRBs discovered to date, the arrival time delay associated
with the dispersion closely follows a $\mathrm{\nu^{-2}}$ frequency
dependence, and the pulse width evolution follows a
$\mathrm{\nu^{-4}}$ frequency dependence for those FRBs where the
signal-to-noise ratio (S/N) has permitted frequency-dependent width
measurements \citep{Thornton}.  Both properties are consistent with
propagation through a sparse, non-relativistic plasma.  

A few years after the publication of the first FRB \citep{Lorimer}
another population of sources (dubbed Perytons) was identified at the
Parkes 64\,m radio telescope that were clearly not of
extra-terrestrial origin.  The Perytons \citep{Burke-spolaor} also
show swept-frequency properties, although they tend to be broader,
mimic interstellar scattering and typically occur during meal-times
onsite.  Unlike the FRBs, the Perytons appeared in all 13 beams of the
Parkes multibeam receiver, indicative of a terrestrial origin. The
Perytons were ultimately shown to be originating from improperly shielded microwave ovens
\citep{Petroff_peryton}. 

Given their large DMs, \cite{Lorimer} and \cite{Thornton} have
proposed that FRBs lie at cosmological distances, and that their DMs
are dominated by propagation through the Intergalactic Medium (IGM),
with minor contributions from the ISM in the Milky Way and the ISM in
a putative host galaxy. Redshift estimates to the sources are made by
subtracting the ISM component of the DM, and ascribing the rest to
propagation through the IGM for which the electron density is
available from cosmological models. For the 15 published FRBs, the
redshift estimates are in the range $0.2 < z < 1.5$, firmly placing
the sources at cosmological distances. As the IGM is thought to
contain 90\% of the Universe's baryons, \citep[e.g][]{Fukugita,
  Savage}, measuring the DMs of FRBs at high redshifts is potentially
a novel way to probe this important cosmological component.  Furthermore, 
if placed at such distances, the unbeamed (isotropic) energies
of the observed FRBs lie in the range $\mathrm{10^{31}}$ to
$\mathrm{10^{33}}$ J \citep{KeanePetroff}. The observed FRBs also have
brightness temperatures well in excess of thermal emission
($\mathrm{T_{b}} > 10^{33}$ K), strongly suggesting coherent emission
\citep{Katz, Luan}.\\
\indent Four events were found by \cite{Thornton} in the high-latitude
component of the High Time Resolution Universe (HTRU) survey at Parkes
\citep{Keith}. From these events, a rate of $\gtrsim
1.0^{+0.6}_{-0.5}\times10^4$ events $\mathrm{sky^{-1}}$
$\mathrm{day^{-1}}$ was estimated. If the redshifts ascribed to the
bursts are valid, the volumetric rate to which this corresponds is
$\sim 2 \times 10^4$ events $\mathrm{Gpc^{-3} yr^{-1}}$, which is
similar to the volumetric rate for Soft Gamma-ray Repeaters (SGRs) ($<
2.5 \times 10^4$ events $\mathrm{Gpc^{-3} yr^{-1}}$), and within an
order of magnitude of the volumetric rate of core collapse (Type II)
supernovae ($\sim 2 \times 10^5$ events $\mathrm{Gpc^{-3} yr^{-1}}$)
\citep{Kulkarni}.

A cosmological origin for the excess DM of FRBs is not the
only possibility, as FRBs could be Galactic objects in high electron
density environments which electron density models for the Milky Way
do not capture. This has been discussed by \citep{Loeb}, who propose
FRBs originate from low mass main sequence ``flare stars''. No
consensus has emerged regarding the progenitors of FRBs no matter
whether Galactic or extra-Galactic, with possibilities including
flare stars \citep{Loeb} (Galactic) and extra-Galactic sources such
as annihilating blackholes \citep{Keaneburst}, giant flares from
soft gamma-ray repeaters \citep{Popov, Thornton, Lyubarsky}, binary
white dwarf mergers \citep{Kashiyama}, neutron star mergers
\citep{Totani}, collapsing supramassive neutron stars
\citep{Falcke}, radio emission from pulsar companions \citep{Zarka},
dark matter induced collapse of neutron stars \citep{Fuller} and the
radio emission from pulsars \citep{Wasserman, Connor}. In this paper
we concentrate explicitly on an extra-Galactic origin for FRBs.


\indent We present here simulations of a cosmological population of FRBs,
under assumptions about their energy distribution, their spatial
density as a function of redshift and the properties of the ISM and
IGM (Section \ref{sec:MC}), finding they are broadly consistent with
origin at cosmological distances. The analysis of the models and the
results are discussed in Section \ref{sec:AR}, in comparison with data
from the HTRU survey. We present log$N$-log$\mathcal{F}$ curves and
discuss the FRB rates at Parkes and UTMOST in Section \ref{sec:AS} and
finally our summary and conclusions in Section \ref{sec:CON}.


\section{Monte Carlo Simulations}
\label{sec:MC}

\begin{figure*}
\centering
\includegraphics[width=6 in]{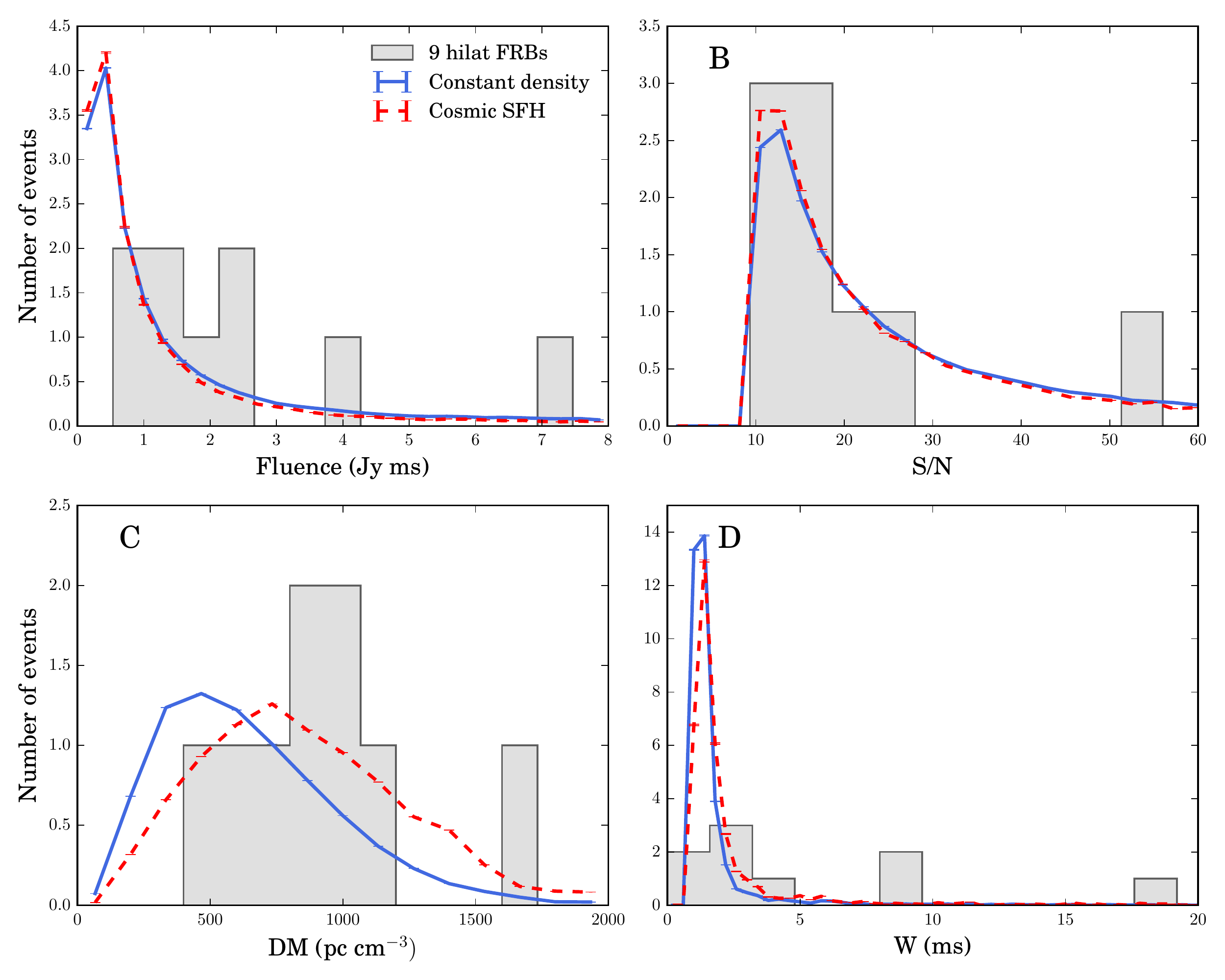}
\caption{Simulated and observed distributions of fluence, S/N, DM and
  width for the 9 Parkes events.  The dashed and solid curves
  represent the cosmic SFH and constant co-moving density
  respectively.  The 9 observed FRB events are represented by the
  histograms. The values of the data have been obtained using the
  \textsc{heimdall}$^{\ref{FootNoteForFigureCaption}}$ single pulse
  detection software package.  {\bf Panel A}: Fluence distribution
  predicted by both models. {\bf Panel B}: S/N distribution
  distribution above the detection threshold of the FRBs. {\bf Panel
    C}: FRB distribution as a function of total DM.  {\bf Panel D}:
  The observed widths distribution predicted by both models.
}
 \label{fig:parkes_observed}
\end{figure*}

\begin{figure*}
\centering
\includegraphics[width=6 in]{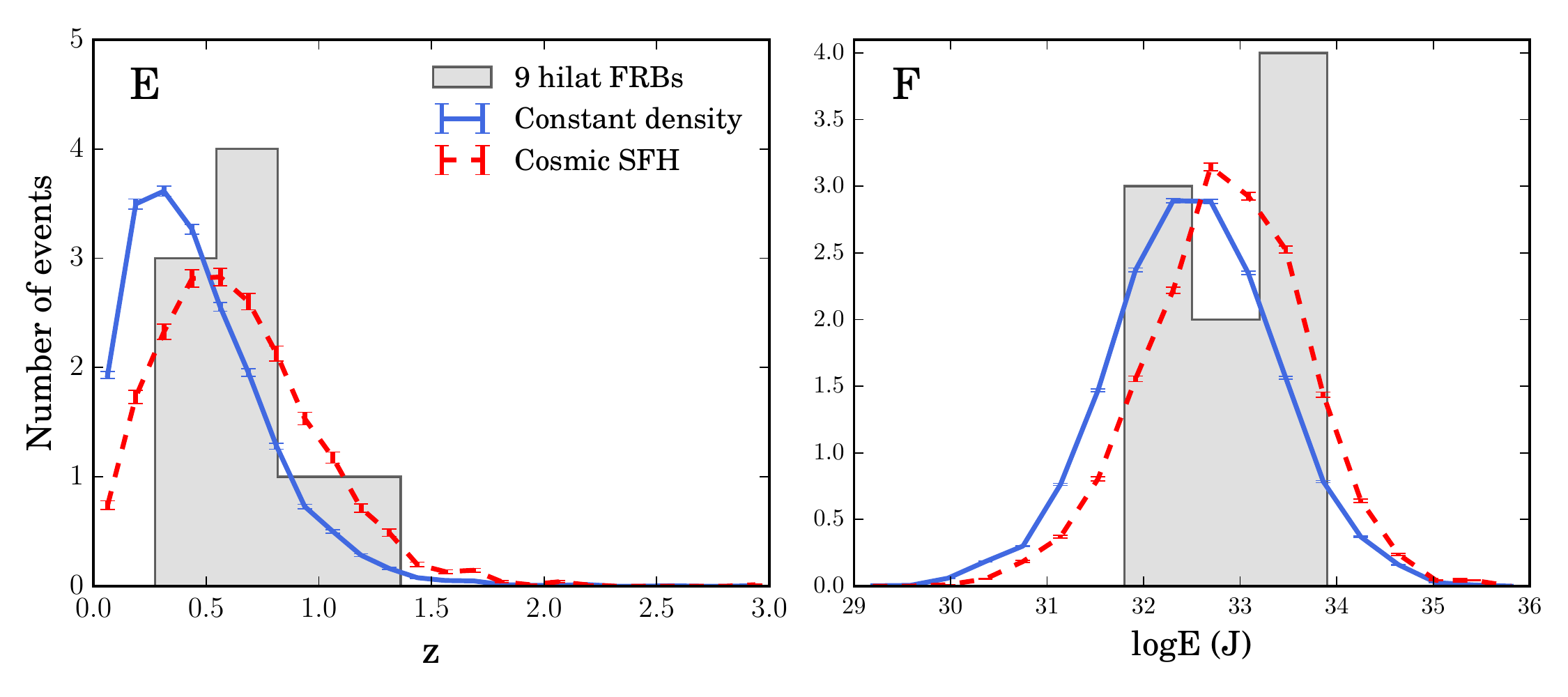}
\caption{Derived FRB parameters from the Monte Carlo simulations of
  FRBs detected in the Hilat survey at Parkes. The dashed and solid
  curves represent the cosmic SFH and constant co-moving density
  models for the FRB spatial densities respectively. The 9 observed
  Parkes FRB events are represented by the histograms. {\bf Panel E}:
  FRB distribution as a function of redshift. {\bf Panel F}: Unbeamed
  energy distribution of the FRBs.}
\label{fig:parkes_inferred}
\end{figure*}

The High Time Resolution Universe (HTRU) survey at Parkes samples the
transient radio sky with 64 $\upmu$s resolution at 1352 MHz and has a
bandwidth of 340 MHz. The observing band is sub-divided into 390.625
kHz frequency channels. HTRU is composed of three sub-surveys at low,
intermediate and high Galactic latitudes. The simulations in this
paper are of the high latitude (Hilat) region of the survey --- 34099,
270-sec pointings at declinations $\delta$ $< 10^\circ$ --- where 9 of
the 18 known FRBs have been discovered (\citeauthor{Thornton}
\citeyear{Thornton}; \citeauthor{Champion} \citeyear{Champion}), and
of the intermediate latitude (Medlat) region, which yielded no FRBs
\citep{Petroff}. \cite{Petroff} and \cite{Hassall} have carried out
studies similar to ours, to model the detectability of FRBs using
simulations and analytic methods respectively. \cite{Petroff}
simulated the effects of dispersion smearing which is the pulse
broadening caused by the adopted frequency resolution, interstellar
scattering and sky temperature on FRB sensitivity at Parkes, in the
Medlat region.  \cite{Hassall} used analytical methods to derive the
detection rates at various telescopes operating over a wide range of
frequencies. Our simulations are of FRB events at cosmological
distances under assumptions about their co-moving density with
redshift, and include the effects of ISM scattering, IGM scattering,
dispersion smearing, sky temperature and telescope beam pattern. We
produce estimates of the energy, fluence, signal-to-noise ratio (S/N),
pulse width, DM and redshift distributions for FRBs with our models,
and compare them to the 9 FRBs detected in Hilat. We
perform two classes of simulations:

\begin{enumerate}
\item in section \ref{sec:MC} we generate numerous events such that
the Poisson noise of the simulations in Figures
\ref{fig:parkes_observed} and \ref{fig:parkes_inferred} is
negligible compared to the noise of the 9 hilat events,

\item in section \ref{sec:MH} we generate thousands of short runs with
an average of 9 events per simulation to estimate and compare the
slopes of their log$N$-log$\mathcal{F}$ curves with the slope of the
log$N$-log$\mathcal{F}$ curve of the 9 hilat FRBs.
\end{enumerate}

For simplicity, FRB events in our simulations are assumed to be
radiating isotropically at the source with a flat spectrum to be
consistent with what is seen at 20 cm with Parkes.  Their intrinsic
energy distribution is assumed to be log-normal. We adopt a
$\Lambda$CDM model with matter density $\Omega_{m}$ = 0.27, vacuum
density $\Omega_\Lambda$ = 0.73 and Hubble constant $H_{0}$ = 71 km
$\mathrm{s^{-1}} \mathrm{Mpc^{-1}}$ \citep{Wright}. The comoving
number density distribution of FRBs in the simulations is assumed to
be either a constant, or proportional to the cosmic star formation
history (SFH). We adopt the SFH from the review paper of
\cite{Hopkins} as typical of cosmic SFH measurements, which show a
rise in the star formation rate of about an order of magnitude between
the present ($z = 0$) and redshifts of $z \sim$ 2 (see their Figure
1). It has the parametric form $\dot{\rho_{*}} =
(a+bz)h/[1+(z/c)^{d}]$ where $h = 0.7$, $a = 0.0170$, $b = 0.13$, $c =
3.3$ and $d = 5.3$ (see their Section 4). We do not
explicitly set the comoving number density of FRBs in the simulation
: we compute the maximum in the product of SFH and comoving volume
of each shell of width $dz$ as a function of $z$, and generate Monte
Carlo events under this function. This allows the simulation to
generate events at the maximum rate, which is important as our run
times can be quite long (c.f. section \ref{sec:AR}).


The total DM for any given FRB is assumed to arise from a component due to
the IGM, a component due to the ISM in a putative host galaxy and a
component due to the ISM of the Milky Way:

\begin{equation} \label{eq:dmtot}
\mathrm{DM_{tot}} = \mathrm{DM_{IGM}} + \mathrm{DM_{ISM}} + \mathrm{DM_{host}}.
\end{equation}

These different DM components are modeled as follows:

\begin{enumerate}

\item the DM due to the IGM is assumed to be related to the redshift
  of the source via the simple scaling relation $\mathrm{DM_{IGM}} =
  1200 z \, \mathrm{pc} \, \mathrm{cm^{-3}}$ with a
  1$\sigma$ scatter of order $\sim20\%$ over the redshift range and DM range 
  of interest (DM$>$100, $0.5\lesssim z \lesssim 2$) \citep{Ioka, Inoue}.

\item The contribution due to the ISM of the Milky Way along the line
  of sight to each event is taken from the NE2001 model of
  \cite{Cordes} which includes the electron density distributions in
  the thin disk, thick disk, spiral arms and Galactic Center
  components. For the high Galactic latitude regions simulated, this
  is generally, $\lesssim  $ 50 pc cm$^{-3}$.

\item The DM contribution of a putative host galaxy will depend on
  galaxy type, the FRB site within it and the viewing angle.
  {\cite{Xu} have modeled the DM distributions due to the ISM for FRBs
    arising in elliptical, dwarf and spiral galaxies. They scale the
    NE2001 model of the Milky Way ISM to the integrated
    $\mathrm{H}{\alpha}$ intensity maps for such hosts, to represent
    their electron density distributions. The ensemble average DM
    distribution for dwarf galaxies is 45 $\mathrm{pc} \,
    \mathrm{cm^{-3}}$ and for elliptical galaxies is 37 $\mathrm{pc}
    \, \mathrm{cm^{-3}}$}. For spirals, they derive the weighted
  average of the DM distribution over a range of inclination angles
  ($0^{\circ}, 30^{\circ}, 60^{\circ}, 75^{\circ}, 90^{\circ}$) to be
  142 $\mathrm{pc} \, \mathrm{cm^{-3}}$. Noting that
    there may be more than one type of FRB progenitor \cite{Masui}
    conclude that their particular FRB could have occurred in a high
    density or star forming region of a host galaxy due to its high
    linear polarisation. Observationally, the galaxy stellar mass
    function distribution peaks near the Milky Way mass \citep{Robles}
    (their figure 9), and we assume the DM properties of the Milky Way
    are typical of a host FRB galaxy. Probing many random lines of
    sight through the NE2001 model, we derive a median DM of $\sim 70$
    $\mathrm{pc} \, \mathrm{cm^{-3}}$ for the Milky Way. Given the
    wide range of DM estimates above, and the uncertainty even as to
    what typical host galaxies are and the sites of FRBs within them,
    we have decided to follow \cite{Thornton} and \cite{Xu}, and
    assume a DM value of $\mathrm{DM_{host}} \sim100 \, \mathrm{pc} \,
    \mathrm{cm^{-3}}$ as typical over a range of hosts and inclination
    angles. This assumption is somewhat \textit{ad hoc}, but does have
    the advantage of facilitating comparison with previous work. The
    assumed DM of the host is a small fraction of the total DM to FRBs
    both in our observed samples and in the simulations, and we could
    vary this host galaxy DM over the full range discussed above ($40
    \la {\mathrm DM} \la 140$) and not affect the conclusions of the
    paper.

\end{enumerate}

\addtocounter{footnote}{1}\footnotetext{http://sourceforge.net/projects/heimdall-astro/
\label{FootNoteForFigureCaption}}

In the simulation, events are generated out to a redshift $z = 3.0$ in
shells of width $dz$ = 0.01, each populated in proportion to the
co-moving volume of the shell and weighted by the star formation rate (SFR) at its redshift
$z$ (in ``SFH'' type models). Events are distributed randomly over the
sky surveyed by Hilat in proportion to the total time spent on sky
(i.e. the product of the number of pointings and the integration time
per pointing). No events are generated north of declination $\delta$ =
+10$^ \circ$, the Northern limit of the survey performed at Parkes.

The fluence $\mathcal{F}$ (in Jy ms) at the telescope is derived from
the energy at the source $E$, the luminosity distance in the
$\Lambda$CDM cosmology and a factor of $(1+z)$ representing the
redshifting of the observed frequency range, given by:

\begin{equation}
\mathcal{F} = \frac {10^{29} \, E } {4 \pi {D_{L}}^{2}(z) \, {B}\, (1+z)} \, \, \mathrm {Jy \,ms}
\end{equation}

\noindent where $z$ is the redshift; $D_{L}(z)$ is the luminosity
distance in pc; $E$ is isotropic emitted energy in J; $\textit{B}$ is
the bandwidth of the receiver system in Hz. The S/N of
  each event is determined using the radiometer equation,

\begin{equation}
\mathrm{S/N} = \beta \, \frac {S \, G \, {\sqrt{B \, t \, N_\mathrm{p}}}} {T_\mathrm{rec} + T_\mathrm{sky}},
\end{equation}

\noindent where ${S}$ is the flux of the signal in Jy, $\beta$ is the
digitisation factor $\simeq$ 1.0, $B$ is the bandwidth in Hz,
${N_\mathrm{p}}$ is the number of polarisations, $t$ is the pulse
width in seconds, ${T_\mathrm{rec}}$ and ${T_\mathrm{sky}}$ are the
receiver and sky temperatures in K respectively, and $G$ is the system
gain in K Jy$^{-1}$. 

Additional simulations of the FRB rates in other surveys are made
later in the paper, and the parameters adopted in those simulations
are shown in Table \ref{tab:specs}.

The brightest FRB in \cite{Thornton}, namely FRB110220 was detected
with S/N of $\sim$ 50 and has an estimated energy $E$ = 10$^{32.5}$ J
at source, a pulse width of $W$ = 5.6 ms, redshift of $z = 0.81$ and a
luminosity distance of $D_{L}(z) = 5.1$ Gpc. \cite{Thornton} assumed
the FRBs were radiating into 1 steradian (that is with a beaming
fraction of 1/4$\pi$), whereas we assume isotropic radiation instead
for simplicity. Accounting for this factor means that the isotropic
energy of FRB110220 in the rest-frame is $E$ = 10$^{33.6}$ J and its
fluence is 7.3 Jy ms.

\subsection{Scattering}

Along the path from source to receiver, a radio pulse may be broadened
in several ways. We assume the scatter-broadening time ($\tau$) of a
pulsed signal passing through the ISM is related to the DM by the
empirical function derived by \cite{Bhat}:

\begin{equation} \label{eq:ISM}
\begin{split}
\textrm{log}(\tau_{\mathrm{ISM}}) = -6.5 + 0.15 \, \textrm{log($\mathrm{DM_{ISM}}$)} \\
+ 1.1 \,\textrm{log($\mathrm{DM_{ISM}}$)}^{2} - 3.9\,\textrm{log}{\nu}
\end{split}
\end{equation}

\noindent where $\tau_\mathrm{ISM}$ is in ms and $\nu$ is in GHz. Rescaling
the scatter-broadening time through the ISM for the IGM,
\cite{Lorimer2013} arrived at an upper limit to the average amount of
scattering as a function of DM, with the scattering due to the IGM
being 3 orders of magnitude smaller than that due to the ISM, i.e.

\begin{equation} \label{eq:IGM}
\textrm{log}(\tau_\mathrm{IGM}) = \textrm{log}(\tau_\mathrm{ISM}) - 3.0.
\end{equation}

\noindent This rescaling on
  scattering in the IGM is still consistent with the observed
  widths of the majority of the FRBs discovered to date
  \citep{Lorimer2013}.


Additionally, the pulse is broadened or smeared across frequency channels
because of the adopted frequency resolution ${\tau_\mathrm{DM} = 8.3
  \,\Delta\nu \, \mathrm{DM} \,\nu^{-3}\, (\mathrm{\mu\, s}})$ where
DM is in $\mathrm{pc\, cm^{-3}}$, $\Delta\nu$ is the channel bandwidth
in MHz and $\nu$ is in GHz. The observed width $W$ of the FRB taking
into account the different contributing components is:

\begin{equation} \label{eq:width}
W^2 = \mathrm{\tau^2_{IGM}} + \tau^2_{\mathrm{ISM}} +
\tau^2_{\mathrm{int}} + \tau^2_{\mathrm{DM}} + \tau^2_{\mathrm{\delta
    DM}} + \tau^2_{\mathrm{samp}} + \tau^2_{\mathrm{\delta\nu}},
\end{equation}

\noindent where the first two components are the scattering times due
to the IGM and ISM, $\tau_{\mathrm{int}}$ is the (unknown) intrinsic
width of the pulse, $\tau_{\mathrm{DM}}$ is due to the DM smearing,
$\tau_{\mathrm{\delta DM}}$ is the second order correction to the DM
smearing, $\tau_{\mathrm{samp}}$ is due to the adopted sampling time
and $\tau_{\mathrm{\delta\nu}}$ is is the filter response of an
individual frequency channel \citep{Cordes2003}. The
$\tau_{\mathrm{\delta DM}}$ and $\tau_{\mathrm{\delta\nu}}$ terms are
typically negligible in the context of our modelling. For the FRBs
discovered at Parkes to date, $\tau_{IGM}$ ranges from $\sim$2
$\upmu$s to $\sim$40 ms and $\tau_{ISM}$ ranges from $\sim$40 ns to
$\sim$10 ms. Previous studies dealing with FRB detectability have
assumed either a ``no scattering" scenario or a strong ISM-like
scattering scenario for the IGM, as its properties are highly
uncertain. \cite{Macquart} have suggested that if the latter scenario
was true, the FRB pulses will be rendered undetectable at current
telescopes, concluding that the IGM scattering was likely weak
($\leqslant 1\, \mathrm{ms}$). We may therefore be sampling a highly-selected
population of FRBs, both in terms of luminosity and scattering. 

The total width of a simulated event $W$ affects its S/N ratio,
scaling it down by a factor proportional to $\sqrt{W}$. This
essentially limits the horizon of the HTRU survey to $z \sim 2$ as
dispersive effects beyond this redshift rapidly degrade the S/N of
even the brightest events to well below the adopted threshold of 10.
Consequently, we use $z = 3.0$ as the high redshift cut-off in the
simulations. This is sufficiently far to sample the dispersion measure
space of the known FRBs.

\subsection{Measured signal-to-noise ratios}

The sky temperature additionally degrades the S/N particularly for sources close to the Galactic plane. We adopt a receiver
temperature\footnote{www.parkes.atnf.csiro.au/observing/documentation/user$\_$guide}
of ${T_\mathrm{rec}}$ = 21 K at Parkes and estimate the sky
temperature (${T_\mathrm{sky}}$) at the Galactic longitude and
latitude $(l,b)$ of the source from \cite{Haslam} who mapped the sky
temperature at 408 MHz with a resolution of $0.85^{\circ} \times
0.85^{\circ}$.  We scaled the survey frequency of 408 MHz to the HTRU
frequency of 1.4 GHz by adopting a spectral index of $-$2.6 for the
Galactic emission \citep{Reich}, i.e.,

\begin{equation}
T_\mathrm{sky} = T_\mathrm{sky_{(l,b)}} \bigg(\frac{\nu}{408.0 \, \mathrm{MHz}}\bigg)^{-2.6}.
\end{equation}

\noindent The S/N of each FRB event is then reduced by the additional
factor ${T_\mathrm{rec}}/({T_\mathrm{rec}}+{T_\mathrm{sky}})$. For
most sources this is a negligible correction, becoming important only
near the Galactic centre and low in the Galactic
plane. $T_\mathrm{sky}$ at high latitudes is typically
$\sim 1$ K and lies between 3 and 30 K over the intermediate
latitude regions.

The S/N is finally degraded depending on a randomly chosen position in
the beam pattern. For Parkes, each beam of the multi-beam receiver is
represented as an Airy disk with a 14.4 arc-minute full-width
half-maximum.  It should be noted that the effect of the beam pattern
is quite significant on the distribution of both the event S/N and the
apparent luminosity; this is discussed in detail in Section
\ref{sec:lognlogf}.



\section{Analysis and Results}
\label{sec:AR}

We have simulated FRBs in two models for their co-moving number density
(either following the SFH, or simply constant density) and either
including or excluding the effects of scattering.  In each model, we
adopt a log-normal source luminosity distribution, centered on a mean
energy $E_{0}$ and spread $\sigma_\mathrm{logE}$. The average energy of the 4
\cite{Thornton} events in \cite{KeanePetroff} correcting for the beaming fraction is
$10^{32.8}$ J. We initially adopt log\,$E_{0} = 32.8$ and a spread of
$\sigma_\mathrm{logE} = 1.0$, as this is the order of magnitude scatter seen on
the \cite{Thornton} events. Within a given model choice
for the source density with redshift, $E_{0}$ and $\sigma_\mathrm{logE}$ are
the 2 free parameters.

The simulations were run on 12 CPU cores with runtimes of a few days
on the gSTAR national facility at the Swinburne University of
Technology. Millions of FRBs are typically generated in the runs, the
vast majority of which are too dim to see. We ran the simulations
until we had $\sim$5000 FRBs that passed the selection
criteria, to ensure good statistical sampling. The distributed
properties of these FRBs are normalized and compared to the 9 observed
Hilat events. 

Slightly different selection criteria have been used by
various authors to find FRBs. \cite{Thornton} used S/N$>$9 and
DM$>$100 $\mathrm{pc\,cm^{-3}}$ and \cite{Champion} use
the same selection criteria as \cite{Petroff}, notably
S/N$\geqslant$10, DM$\geqslant$ 0.9 $\times$ DM$_\mathrm{MW}$ and
$W$$\leqslant$16.3 ms. We use the criteria S/N$\geqslant$10 and
$W$$\leqslant$32.786 ms for the selection of the candidates in the
simulations. We adopt an upper limit of 32.786 ms for the width
motivated by the fact that the broadest FRB discovered in Hilat has a width of
$\sim$19 ms, and in any case broader events still would have to be
extremely bright to have S/N$>$10. We do not apply a DM
threshold for the Hilat region as we are only sensitive to DM$>$100
$\mathrm{pc\,cm^{-3}}$ in keeping with \cite{Thornton}, due to
assuming the value of DM$_\mathrm{host}$ to be 100
$\mathrm{pc\,cm^{-3}}$. Tests showed that the differences in these
selection criteria are minor and have negligible effect on our basic
results. In particular every observed FRB fits each of these criteria.


\subsection{Monte Carlo results for Parkes}
\label{sec:SIM}

Figures \ref{fig:parkes_observed} and \ref{fig:parkes_inferred}
display the results of the simulations of the cosmic SFH
($\rho_\mathrm{\text{\tiny{FRB}}}(z)$ =
$\rho_\mathrm{\text{\tiny{SFH}}}(z)$) and constant-density
($\rho_\mathrm{\text{\tiny{FRB}}}(z)$ = constant) models with
scattering included, as seen by Parkes, overlaid on histograms of the
9 observed Hilat FRBs (\citeauthor{Thornton} \citeyear{Thornton};
\citeauthor{Champion} \citeyear{Champion}).  Figure \ref{fig:parkes_observed} shows
observational parameters for each burst and Figure
\ref{fig:parkes_inferred} displays the parameters that are derived.
We quantify the goodness of fit of the model to the observations in
Section \ref{sec:SA}.

The fluence distribution of our simulated events is displayed in panel
A of Figure \ref{fig:parkes_observed} and their S/N distribution in
panel B, each compared to the 9 Hilat events. All the observed Hilat
FRBs have fluences between 0.7 and 7 Jy\,ms. Both models peak at
$\sim$ 0.5 Jy\,ms and are reasonable matches to the observations. The
S/N distributions of both models contain a large number of events just
above the detection threshold of 10 and then gradually decline towards
higher values; both appear to agree with the observations reasonably. Panel C
shows the DM distribution of the models and the observations: this is
similar to the panel showing the redshift distribution, since they are
closely related. Both the cosmic SFH and constant density models are
in agreement with the observed data. 

The width of an FRB pulse affects its detection S/N. In the observer's
rest frame, the width results from the sum of contributions from
scattering due to the ISM and IGM \citep[Equations \ref{eq:ISM} and
  \ref{eq:IGM}, see][]{Bhat, Lorimer2013} DM smearing time and the
intrinsic width.  Panel D of Figure \ref{fig:parkes_observed} displays
the distributions of the observed widths of the sources. We found
neither model to agree with the data very well and may be a result of our
simplistic model of intergalactic scattering discussed below.

The adopted model for the spatial density of the sources in Figure
\ref{fig:parkes_inferred} panel E does not have much effect on their
redshift distribution, with only a small excess of sources produced at
$0 < z < 0.5$ for the constant density model compared to the cosmic
SFH model. As expected, we see a tendency for more events at higher
redshift for the SFH model compared to the constant density
model. Panel F shows the energy distribution (at source and in-band)
of the FRBs and the models. Both models are only sensitive to the
bright tail of the adopted log-normal energy distribution function and
have similar mean values to that of the 9 observed FRBs. Since the
mean energy $E_{0}$ of the adopted luminosity function is a free
parameter we adjust this to achieve good fits to the observed
luminosities in panel F of Figure
\ref{fig:parkes_observed}. Acceptable fits are obtained for both
models by adopting $E_{0} = 10^{31.2}$ J, with a log normal-scatter of
$\sigma_\mathrm{logE} = 1.0$. This adopted luminosity function is a parameterised luminosity function only and
does not possess any physical significance.

\subsection{Statistical analysis}
\label{sec:SA}

Kolmogorov-Smirnov tests (K-S) were performed on all the distributions
in Figures \ref{fig:parkes_observed} and \ref{fig:parkes_inferred} and
the resulting probability statistics $p$ are given in Table
\ref{tab:p-values}. A $p$-value of $<0.05$ is our
  criterion for deciding if the two distributions differ. Each model
  was compared against the data for the 9 FRBs.


The $p$-values show that both models are consistent with the observed
distributions of redshift, energy, fluence S/N and DM but, as already
noted above, we have difficulty modelling the effect of scattering on
the FRBs. The $p$-values of 0.013 (density of FRBs proportional to the
cosmic SFH with redshift) and 0.001 (density of FRBs constant with
redshift) reject the hypotheses that both models and the FRB data are
from the same population. The present sample of 9 events is thus
insufficient to distinguish between these models per se (the poor
match to the distribution of pulse widths in both models
notwithstanding).
For FRBs discovered at Parkes, our simulations indicate that of order
50 FRBs are required to distinguish between the two FRB number density
models at the 95$\%$ confidence level. This certainly highlights the
need to discover FRBs more efficiently, as the present discovery rate
is only of order 1 per 12 days on sky at Parkes.

\begin{figure}
\centering
\includegraphics[width=3.1 in]{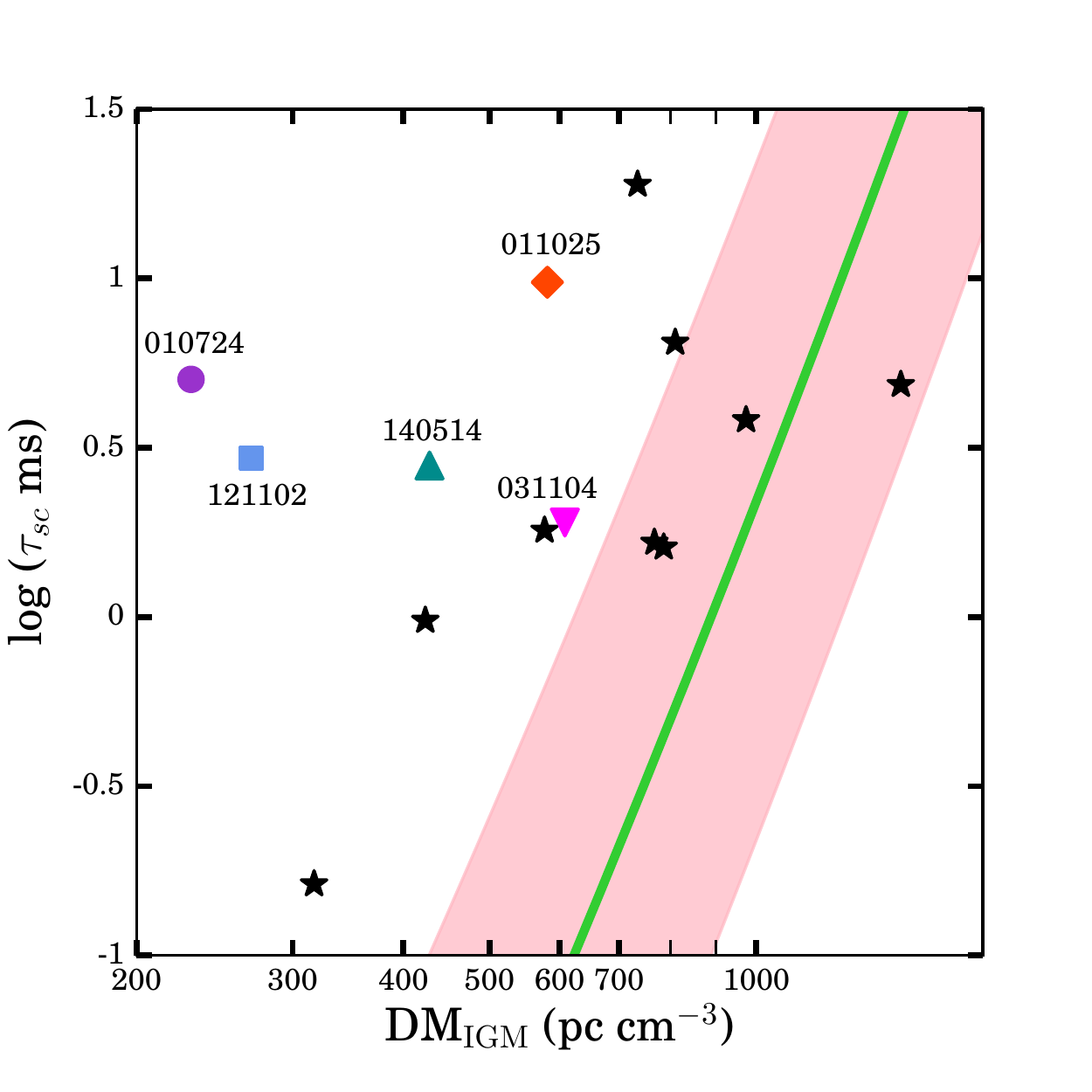}
\caption{Adopted model of the scattering time due to the
    IGM (solid) at 1.4 GHz for Parkes versus estimated dispersion
    measures. Stars represent the 9 Hilat events and other markers
    represent FRBs discovered in various other surveys. The shaded
    region around the fitted line to the equation represents the order
    of magnitude scatter adopted in the simulation. Note that one
    Hilat FRB lies below the IGM scattering relation (see Equation \ref{eq:IGM})
    but is still within the
    adopted 1-sigma spread.}
\label{fig:scatter}
\end{figure}

To better understand the effective widths, the 14 FRBs at Parkes as a
function of scattering time is shown in Figure \ref{fig:scatter}. The
estimated widths of the events due to IGM scattering and a
possible intrinsic width ($\mathrm{\tau^2_{IGM}} +
\tau^2_{\mathrm{int}}$ = $W^2 - \tau^2_{\mathrm{DM}} -
\tau^2_{\mathrm{ISM}}$ from Equation \ref{eq:width}) are plotted
against our estimate of the contribution to the total DM due to the
IGM alone ($\mathrm{DM_{IGM}} = \mathrm{DM_{tot}} -
\mathrm{DM_{ISM}} - \mathrm{DM_{host}}$ from Equation
\ref{eq:dmtot}).  We see that the scattering times are inconsistent
with Equation \ref{eq:IGM}, and show considerable scatter around
it.

\begin{table}
\centering
\caption{K-S test results for the model distributions against data in
  Figures \ref{fig:parkes_observed} and \ref{fig:parkes_inferred}}
\label{tab:p-values}
\begin{tabular}{ccc}
\hline\hline
\multirow{2}{*}{Parameter} & \multicolumn{2}{c}{$p$-value} \\
                           & Cosmic SFH    & Constant density    \\
\hline
Redshift                   & 0.543         & 0.048               \\
Energy                     & 0.884         & 0.186               \\
Fluence                    & 0.047         & 0.106               \\
S/N                        & 0.258         & 0.078               \\
DM                         & 0.730         & 0.053               \\
Effective width            & 0.013         & 0.001               \\ [1ex] 
\hline 
\end{tabular}
\end{table}

This result highlights the basic difficulty with the IGM model,
apparent in the data (Figure \ref{fig:scatter}), that the pulse widths
of the observed FRBs scatter around the adopted functional form for
the IGM (Equation \ref{eq:IGM}). This behaviour is also seen for
pulsars being scattered by the ISM, for which there is at least an
order of magnitude scatter in the data around the observed pulse width
trend \citep[Equation \ref{eq:ISM}, see][]{Bhat}.  If we assume that
there is a similar scatter around $\tau_\mathrm{{IGM}}$ of an order of
magnitude, we still do not acquire satisfactory fits to the data
within 2$\sigma$ confidence. This suggests that the scattering is not
due to a line-of-sight dependent inhomogeneous IGM. It may be due to
interaction with the ISM of an intervening galaxy or an intracluster
medium along the line-of-sight, although the probability of
intersection at the redshifts modelled is quite low and only a small
fraction of lines of sight may be affected \citep{Macquart}. We have
not attempted to model such effects: our aim is to test a much simpler
model before adding in difficult to test assumptions about the
properties of the IGM.

If we assume that our simulated events have a mean intrinsic width of 3 ms (with
a standard deviation of 3 ms, truncated at 0 ms), the resulting width distributions are found
to be in good agreement with the observed 9 hilat FRBs. The intrinsic
width assumption is motivated by FRB121002 and FRB130729 \citep{Champion} 
which have hints of double, rather than single peaked
pulse profiles. This is a rather \textit{ad hoc} assumption and
further work on this is required once the population is expanded.
The disagreement of the distribution of event widths with the observations 
is the weakest point in our modelling. Clearly, there is a need for more FRBs to resolve
this problem. 


\subsection{The log$N$-log$\mathcal{F}$ of the Hilat events}
\label{sec:lognlogf}

In a Euclidean Universe populated with events (or objects) of fixed
luminosity (i.e. standard candles) and uniform number density, the
number $N$ detected above some flux limit $S$ varies as $N
\propto S^{\alpha}$, where $\alpha=-3/2$. In our model, the
FRBs have a very broad luminosity distribution and are sufficiently
distant that non-Euclidean effects are important. Consequently we do
not expect to see $\alpha = -3/2$.

The very wide range of luminosities of the observed events suggests
they are not particularly good standard candles, and until we have a
redshift of an FRB host galaxy, or some other independent distance
indicator for an FRB, their luminosities are highly dependent on the
assumption that DM is a proxy for redshift. 
The luminosities are dependent on each line of sight being equal to the 
average line of sight in a $\Lambda$CDM Universe. In fact it is the 
small deviations from this that we will use to do some cosmology, 
when we have a lot of FRBs with real redshifts.
In any case, our FRB simulations are for a $\Lambda$CDM cosmology, which affects
$\alpha$. In a $\Lambda$CDM cosmology, $\alpha$ varies smoothly from a
slope of $-3/2$ for the nearby universe, gradually becoming flatter as
further distances are probed. To illustrate, at a redshift of $z \sim
0.7$, typical of FRBs found to date, standard candles yield a
relation with a slope of $\alpha \sim -1$.
There are additional factors which affect $\alpha$. Firstly, the HTRU
survey is ``fluence incomplete'' in the sense that events with the
same fluence are easier to detect if they have narrower pulse
widths. Secondly, propagation of FRB pulses through the IGM causes the
pulses to broaden, reducing their S/N, so that a S/N selected sample
effectively has a distance horizon beyond which pulses are too
scattered to see. This will flatten the
relation as we probe to dimmer events.

It is possible to select a ``fluence complete'' sample of the FRBs,
and compare these to simulation events selected in the same way, but
this would reduce our sample of 9 events to just 4 events. For a
S/N of 10, the fluence completeness limit for Hilat
is $\sim 2$ Jy\,ms \citep{KeanePetroff}. This is an observational
selection, and affects the slope $\alpha$, of the
relation. It is straightforward to include
this effect in the simulations, however, due to our already small sample 
of events we prefer to compare to the full
set of 9 events selected by S/N, rather than a fluence complete set of
4 events.

\begin{figure}
\centering
\includegraphics[width=3.5 in]{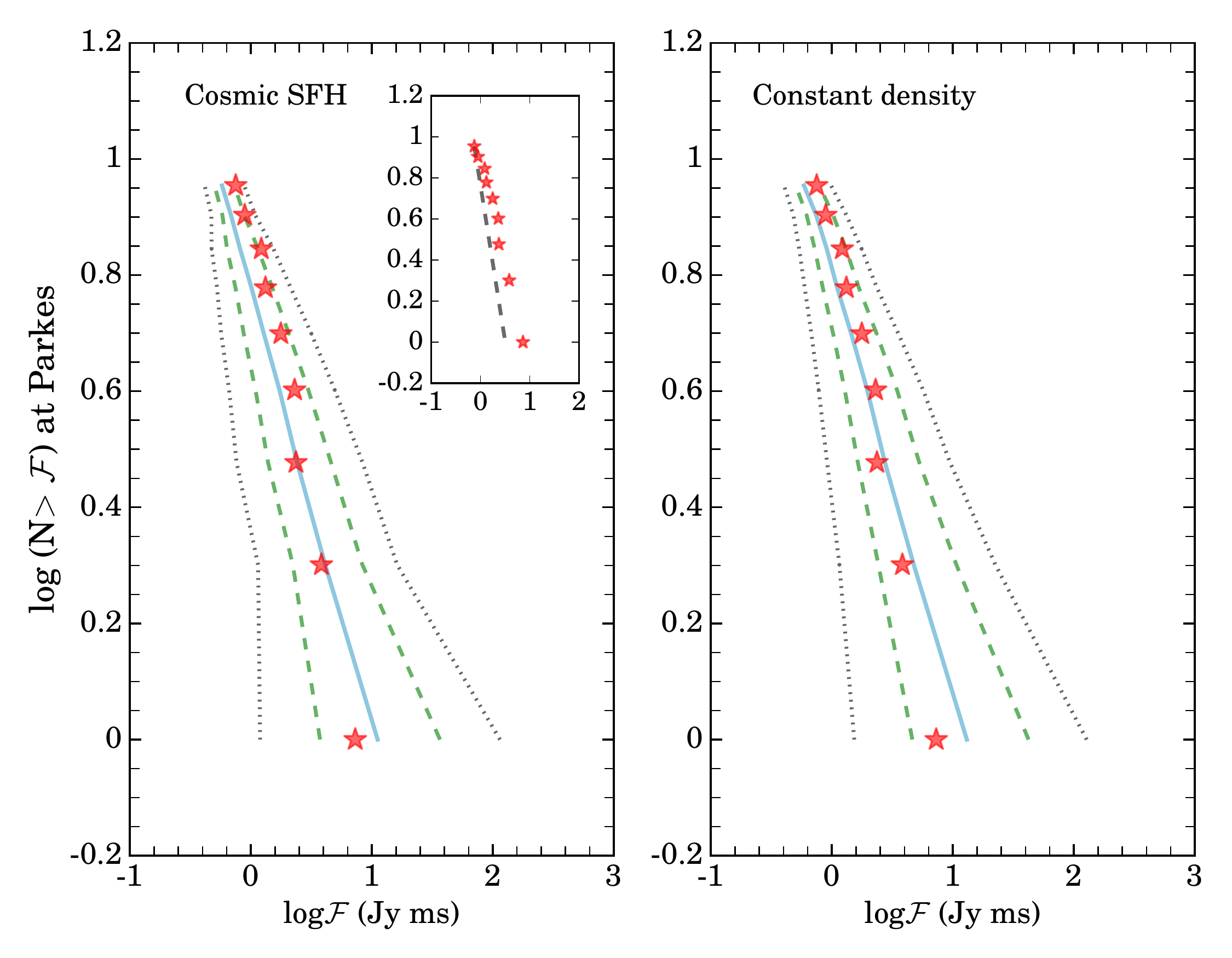}
\caption{The log$N$-log$\mathcal{F}$ curves for the 9 Hilat FRBs and
  the simulation samples.  The left panel displays the cosmic SFH
  \citep{Hopkins} scenario and the right panel displays the constant
  density scenario.  Stars represent the 9 Hilat FRBs and the solid
  line connects the medians of the number densities as a function of
  fluence for the simulation sample. The dashed and dotted
  lines represents the 1$\sigma$ and 2$\sigma$ limits around the
  median for each N. The inset in the left panel exhibits the 9 observed
  FRBs and a fitted slope of $\alpha = -$3/2 for comparison.}
\label{fig:9_events}
\end{figure}

\begin{figure}
\includegraphics[width=8.6 cm]{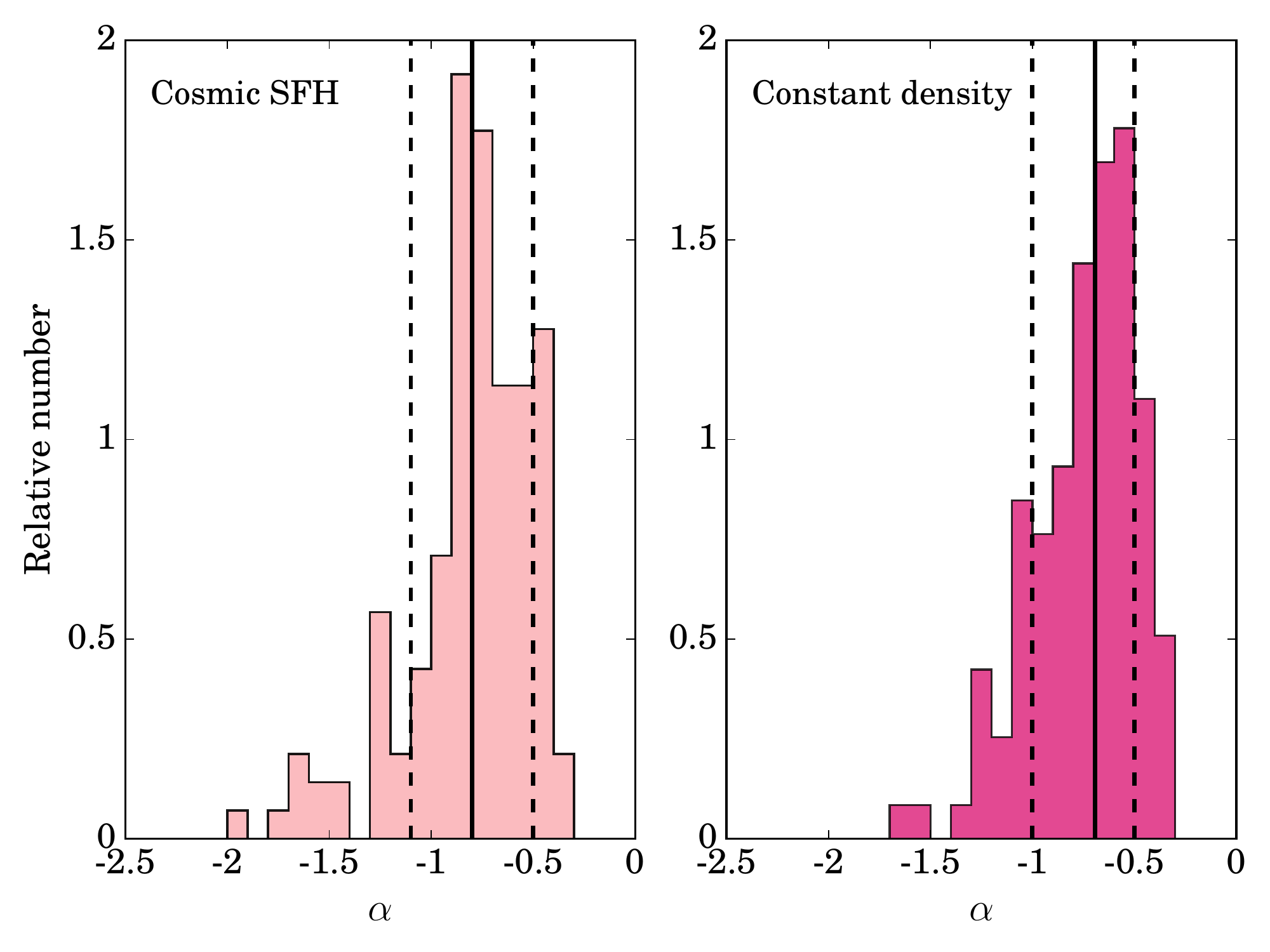}
\caption{The histograms display the slopes $\alpha$, of the simulation
  samples containing exactly 9 events each. The left panel represents
  the cosmic SFH scenario and the right panel represents the constant
  density scenario. The medians of the histograms are
  represented by the solid lines and the 1$\sigma$ scatter from the
  median is marked by the dashed lines. The slope of the 9 FRBs
  $\alpha = -0.9\pm0.3$ is found to be consistent with the simulations
  within the uncertainties.}
\label{fig:slope_dist}
\end{figure}

The log$N$-log$\mathcal{F}$ plot of the 9 Hilat events is shown in
Figure \ref{fig:9_events} --- note that we use the fluence
$\mathcal{F}$ in Jy ms (since FRB detections are width dependent) 
for what would normally be flux density $S$ in Jy. The
cumulative log$N$-log$\mathcal{F}$ relation is reasonably linear for
the 9 events, and has a slope of $\alpha = -0.9\pm0.3$. For the
cumulative curve of only 9 events, sample variance is likely to be a
significant factor. We use the simulations of Hilat 
(as described in section \ref{sec:MC} with selection criteria described in
Section \ref{sec:AR}), which were set up to
yield of order 9 events per run to estimate the error on
$\alpha$. Those realisations which had exactly
9 events were used for comparison with the 9 observed Hilat events. We
have fitted slopes ($\alpha$) to these simulated 9 event samples and
show the distribution of $\alpha$ in Figure
\ref{fig:slope_dist}. The typical error on $\alpha$ is
$\pm$0.1 which is the adopted bin size in Figure
\ref{fig:slope_dist}. The median slope obtained is $\alpha = -0.8$
for the SFH case and $\alpha = -0.7$ for the constant density case,
but with significant scatters (the 1 $\sigma$ limits are shown as
dashed lines) of order $\pm0.3$ for the SFH and $\pm0.2$ for the
constant density around the mean.  Our observed slope of $\alpha =
-0.9\pm0.3$ is consistent with both models.

We conclude that the slope of the log$N$-log$\mathcal{F}$
relation of the 9 observed events is consistent to within the
uncertainties of both the simulated models, indicating that our
measured log$N$-log$\mathcal{F}$ slope is consistent with FRBs being
of cosmological origin. This is in agreement with the 
conclusion of \cite{Katz3} that the log$N$-log$S$ and N vs.
DM distributions are consistent (except for the anomalously bright Lorimer burst) 
with cosmological distances inferred from their DM in a simple
approximation to standard cosmology.

\subsection{Medlat vs Hilat}
\label{sec:MH}

The intermediate-latitude component of the HTRU survey consists of
540-sec pointings in the range $-120^{\circ} < l < 30^{\circ}$ and
$|b| < 15^\circ$.  \cite{Petroff} found no FRBs in this region of the
survey.  Under the assumption that FRBs are isotropically distributed,
scaling from Hilat, and accounting for a slight reduction in their
detectable source density in the Medlat region due to the smearing
effects of the ISM, they estimate the probability of this occurring by
chance as only of order 0.5\%. We simulate both the
Medlat and Hilat regions (adopting 100$\%$ of Hilat and 100$\%$ of
Medlat as the surveyed completeness for the regions for FRBs) to
determine the likelihood of finding zero FRBs in Medlat for 9 discovered
FRBs in Hilat. The simulation for Medlat is otherwise
identical to the one described in Section \ref{sec:MC} except for
the survey parameters i.e. number of pointings, region of sky surveyed, $T_\mathrm{sky}$
corresponding to the region of sky surveyed and
integration time per pointing. The same selection criteria as 
described in Section \ref{sec:AR} are used for selection of
candidates in both Medlat and Hilat.

\begin{figure}
\centering
\includegraphics[width=2.63 in]{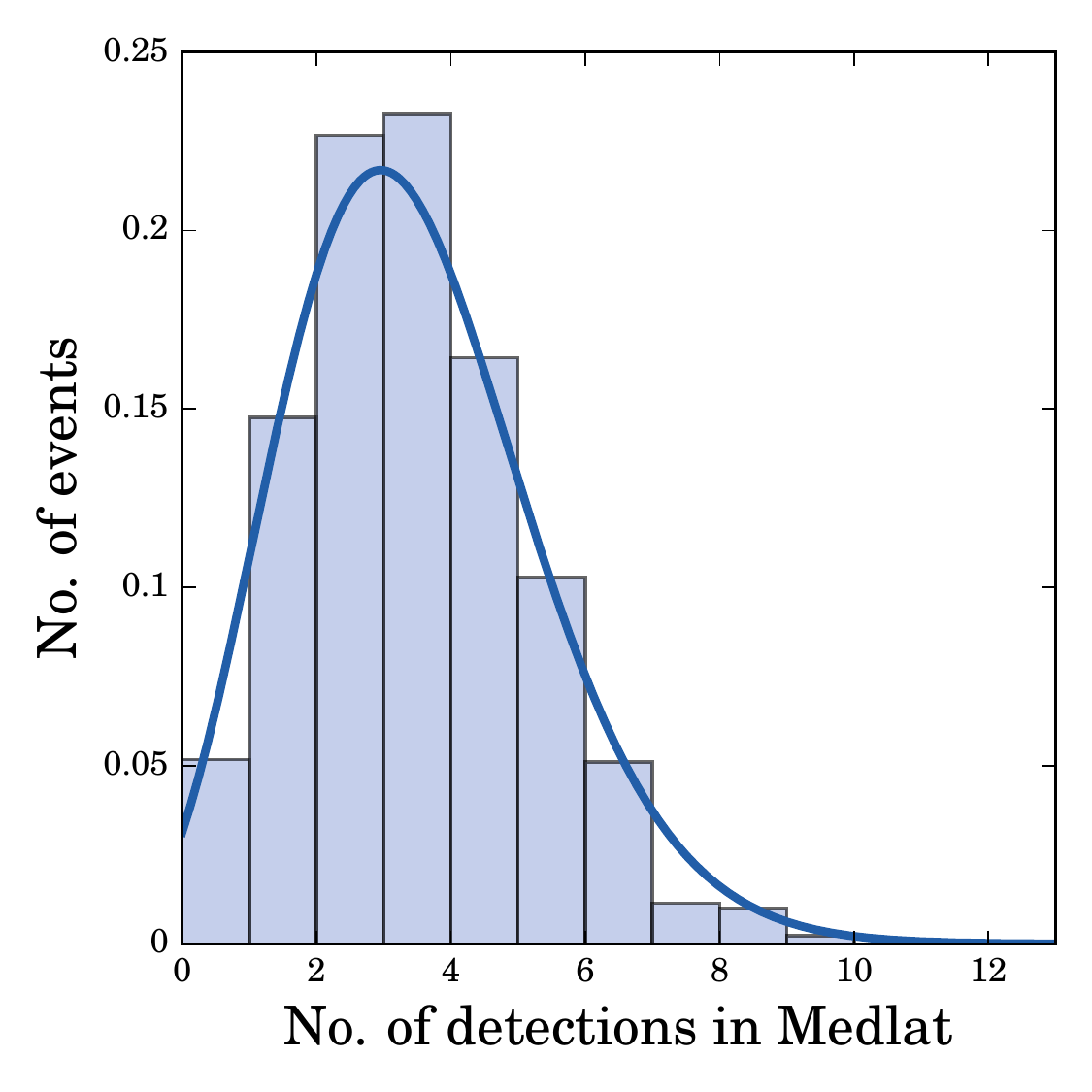}
\caption{Number of FRBs expected in Medlat normalized to the 9 events
  in Hilat. Both surveys are assumed to be fully searched for
  FRBs. The histogram represents the number of FRBs expected in Medlat
  for a corresponding 9 FRBs detected in Hilat. A Poissonian curve is
  fitted to the data. The number of FRBs found in Medlat is zero 5.1\%
  of the time.}
\label{fig:poisson}
\end{figure}

We obtain an average of $\sim3\pm2$ events in our Medlat simulations for every 
9 events in the Hilat simulations, finding
no events just 5.1$\%$ of the time (Figure \ref{fig:poisson}).  The
estimated probability of zero events being seen in Medlat 0.5$\%$ of the
time made by \cite{Petroff} is based on the 4 events detected in the
24$\%$ of the Hilat survey which had been searched at the time.  The
higher probability we estimate of finding no events in Medlat in our
simulations is due to our using the lower all sky rate, now that Hilat
has been completely searched and it only yielded 9 FRBs.

\begin{table*} 
\centering
\begin{minipage}{167mm}
\caption{Specifications of Parkes multibeam, Parkes PAF and UTMOST} 
\label{tab:specs}
\begin{tabular}{c c c c c}
\hline\hline 
Parameter          & Unit                 &   Parkes MB \citep{Keith}  &   Parkes PAF\footnote{http://www.atnf.csiro.au/management/atuc/2013dec/science$\_$meeting/ATUC$\_$PKS$\_$receivers.pdf}   &   UTMOST (Bailes et al. in prep) \\ [0.5ex] 
\hline
Field of View      & $\mathrm{deg^2}$     &   0.55        &   2.2          &   $4.64\times2.14$   \\
Central beam Gain  & K $\mathrm{Jy^{-1}}$ &   0.7         &   0.9          &   3.5 \\
Central beam $T_{sys}$ & K                &   21          &   50           &   70  \\ 
Bandwidth          & MHz                  &   340         &   340          &   16  \\
Frequency          & MHz                  &   1352        &   1352         &   843 \\
Channel width      & MHz                  &   0.390625    & $\sim$1        & 0.78125  \\ 
No. of polarisations  & --                &   2           &   2            & 1   \\
Polarisation feeds    & --                &   Dual linear & Dual linear    & Right circular  \\ [1ex]
\hline  
\end{tabular} 
\end{minipage}
\end{table*}

\begin{table}
\centering
\caption{Minimum detectable flux density for a 10$\sigma$, 1 ms event and
  event rate assuming a Euclidean scaling for the Parkes multibeam, Parkes PAF and UTMOST }
\label{tab:rates}
\begin{tabular}{ccc}
\hline\hline
Telescope/Receiver & $S_\mathrm{min}$ (Jy) & Rate (events $\mathrm{day^{-1}}$)\\
\hline
Parkes MB                  & 0.4        & 0.08$\pm$0.03              \\
Parkes PAF                 & 0.6        & 0.10$\pm$0.04               \\
UTMOST                     & 1.6        & 0.16$\pm$0.06               \\[1ex] 
\hline 
\end{tabular}
\end{table}

\section{The log$N$-log$\mathcal{F}$ of FRB events}
\label{sec:AS}

Our simulations have been used to generate FRB events 
at 2 facilities -- Parkes and UTMOST (Bailes et al. in
prep). UTMOST is the recently upgraded Molonglo Observatory
Synthesis Telescope located about 300 km south-west of Sydney, near
Canberra, and is a field station of the University of Sydney. We
generate events for the specifications of UTMOST and Parkes for the soon to be
installed phased array feed (PAF) receiver in comparison with the
current multibeam receiver (MB) at Parkes. The FRB co-moving density
models, and energy distributions are the same as those described in
Section \ref{sec:MC}. The effective pulse width of each event is
computed using Equation \ref{eq:width}. The S/N of the events were
reduced by a factor of 4 for the events at UTMOST to account for the
fact that it is less sensitive than the MB receiver at Parkes (Caleb
et al. in prep). The Parkes PAF is estimated to have $\sim 50\%$ of
the sensitivity of the
multibeam\footnote{http://www.atnf.csiro.au/management/atuc/2013dec/
  science$\_$meeting/ATUC$\_$PKS$\_$receivers.pdf} which is accounted
likewise. The S/Ns at both telescopes were further reduced by
$\sqrt{W}$ before making the cut-off of S/N$\geqslant$10 and
$W$$\leqslant$32.786 ms.

\begin{figure*}
\centering
\includegraphics[width=6 in]{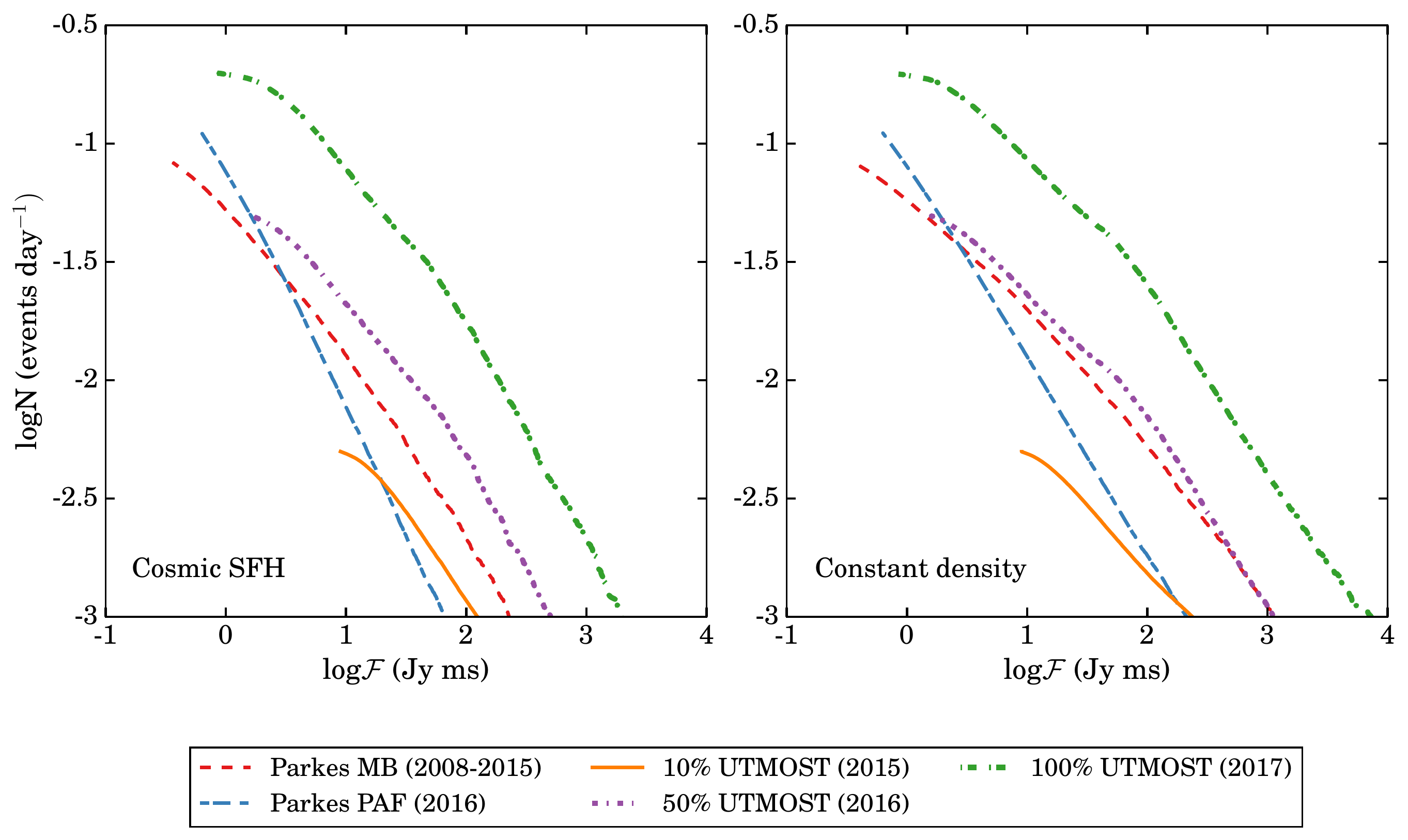}
\caption{The log$N$-log$\mathcal{F}$ curves for different fractional
  sensitivities at UTMOST and MB and PAF receivers at Parkes. The left
  panel displays the log$N$-log$\mathcal{F}$ curves for the cosmic SFH
  \citep{Hopkins} model of the FRB space density with redshift and the
  right panel displays the curves for the constant density model. All
  curves include the ISM and IGM scattering and are normalised to the
  rate of $\sim$ 1 event per 12 days at the Parkes MB and additionally
  to the ratio of their fields of view for UTMOST and the Parkes
  PAF. Uncertainty in the PAF design ssensitivity makes prediction
  difficult, but its wider sky coverage can increase the Parkes
  discovery rate at lower fluences. The fully sensitive UTMOST will
  dominate the event detection rate at all fluences.}
\label{fig:logNlogS}
\end{figure*}

Figure \ref{fig:logNlogS} shows the cumulative log$N$-log$\mathcal{F}$
curves at UTMOST and at Parkes for
both the MB and phased array feed PAF. These curves do not include the
effects of fluence completeness. All curves have been normalised to their respective FRB
rates in Table \ref{tab:rates}, which have been calculated assuming a Euclidean Universe where
the cumulative number density scales as $\propto \mathcal{F}^{\alpha}$
where $\alpha = -3/2$ (Caleb et al. in prep). This is a conservative option, as the slope of
this relation is most likely flatter (as seen in the previous
section), and underestimates the number of events expected.

\section{Discussion and Conclusions}
\label{sec:CON}

We have simulated observational and derived properties of
a cosmologically distributed population of FRBs, for comparison with
the 9 FRBs seen in the HTRU/Hilat survey conducted at Parkes from
2008-2014. Two models for the spatial number density of the FRBs are
examined: firstly, where the co-moving density is a constant,
and secondly, where the number of FRBs is proportional to the cosmic
SFH. The properties of the ISM in the Milky Way and
a putative host galaxy for the FRB are taken into account, and
conservative assumptions are made about the properties of the IGM, the
spectral index of FRBs and their luminosity function. 

The simulated distributions of redshift, energy, DM, S/N, fluence and
effective widths for both the cosmic SFH and constant density models
were compared to the 9 observed FRBs. We achieved reasonable matches
to the data for all these properties except the event widths, by
adjusting only the typical FRB event energy at source (and scatter
around this energy) i.e. by adjusting only their luminosity function.
It proved difficult to fit the distribution of FRB widths
  without making \textit{ad hoc} assumptions about scattering in the
  IGM or the intrinsic widths of the pulses. The simulations are
  intended to look at FRB properties with as simple an assumption set
  as possible; adding in poorly constrained properties as these for
  the FRBs and the IGM for the sake of fitting the pulse widths was
  not pursued. As the pulse widths probe completely different
  properties of FRBs and the IGM, this may prove more fruitful to
  understanding their origin as more FRBs are found. 

The most interesting property of the simulated events is their
distribution of log$N$-log$\mathcal{F}$, where $N$ is the number of
events detected above some fluence $\mathcal{F}$. If the sources have
an even approximately typical luminosity (i.e. are standard
candle-like) then the slope of this relation is a probe of their
spatial distribution. For standard candles of flux $S$ distributed
uniformly in empty, Euclidean space, the slope of the closely related
log$N$-log$S$ relation is well known to be exactly $-3/2$. For FRBs,
the slope of the relation is affected substantially for 3 main
factors: firstly by cosmology (space is non-Euclidean); secondly by
propagation through the IGM (i.e. space is not empty) and thirdly by
selection effects at the telescope (narrower events are detected more
readily than broader ones). A major aim of the simulation is to
quantify these effects.

The observed slope $\alpha$ of the log$N$-log$\mathcal{F}$ of the 9 FRBs
analysed is $\alpha = -0.9 \pm 0.3$. Our simulations are able, in
both scenarios for the number density of the sources with redshift, to
match this slope well, yielding $\alpha = -0.8 \pm 0.3$ for the cosmic SFH and
$\alpha = -0.7 \pm 0.2$ for the constant density case. We conclude
that the properties of the observed FRBs are consistent with arising
from sources at cosmological distances, with the important caveat that
the pulse width distribution does not match our simulation results
particularly well.

The luminosity function of the FRBs is the main free parameter in the
simulations. We adopt a log-normal luminosity function (LF) and adjust
the mean energy $E_{0}$ and spread in energy $\sigma_\mathrm{logE}$.
It is clear from the 9 observed events that a narrow,
standard-candle like LF is an unacceptable fit, since their inferred
intrinsic lumninosities has a spread of about an order of
magnitude.  We measure a mean energy $E_0$ of $\sim 10^{31.2}$ J
with a spread of a factor of 10 in energy. As the observed FRBs very
much sample only the high luminosity tail of this distribution, other
choices for the LF, such as a truncated power law would also
adequately match the data. Our studies show that the
beam pattern of the telescope has a strong effect only when the
number of FRBs is large ($\ga {\mathrm few} \times 100$), which is
then sensitive to the high luminosity tail of events. The LF choice
affects the distributions strongly even for small samples : an LF
with a significant spread in luminosity is required to model the 9
events. Finally, our simulations show that the adopted comoving
density models for the FRBs has weak effects, and large sample sizes
($\ga 100$) are required to probe this.

Future work could implement other LF choices and
  investigate the extent to which the LF and SFH and beam pattern
  affect the observed distributions analysed in this paper: the small
number of FRBs detected to date do not warrant such work here.

Our simulations show that at least 50 FRB events are required to
distinguish, at the 95$\%$ confidence level, between our two tested
models for their cosmological spatial distributions for the specifications
of the Parkes telescope. This argues
strongly for projects to increase the detection rate of FRBs by using
wide field of view instruments, such as UTMOST and CHIME
\citep{Bandura}. Even more important in the immediate future is to
localise events on the sky (to find putative host galaxies for FRBs)
and a number of experiments are ongoing to do this (eg: SUPERB project
at Parkes).

We have applied our simulations to the Medlat survey at Parkes (which
is part of the HTRU survey), which surveyed a lower Galactic latitude
region of the sky with longer integrations. Our simulations of this
survey supports the conclusion of \cite{Petroff} that the sky rate of
FRBs in \cite{Thornton} is overestimated by about 50$\%$, or that FRBs
are not distributed isotropically on the sky.

We simulate FRB rates at two other facilities : at UTMOST (first survey
results of which are in a companion paper Caleb et al. in prep) and at
Parkes with the planned Phased Array Feed (PAF), under conservative
assumptions about the spectral index of FRBs, and the sensitivity of
the instruments. UTMOST has the capability, at full design sensitivity
to dominate the FRB detection rate. Uncertainty in the final PAF
design sensitivity makes prediction difficult, but its wide sky
coverage has the potential to increase the FRB discovery rate of FRBs
close to the fluence limit. The fully sensitive UTMOST will dominate
the event detection rate at all fluences.


\section*{Acknowledgements}
The authors gratefully acknowledge valuable discussions with Emily
Petroff, J-P Macquart and Alan Duffy. This work used the gSTAR national
facility which is funded by Swinburne and the Australian Governments
Education Investment Fund. Parts of this research were conducted by
the Australian Research Council Centre for All-Sky Astrophysics
(CAASTRO), through project number CE110001020.


\bibliographystyle{mnras}

\bibliography{Arxiv}


\end{document}